\documentclass[runningheads]{llncs}
\usepackage[T1]{fontenc}
\usepackage{amsmath}
\usepackage{amssymb}
\usepackage{booktabs}
\usepackage{multirow}
\usepackage{tabularx}
\usepackage{graphicx}
\usepackage{hyperref}
\newcommand{\Framework}{$\mathcal{F}$}

\usepackage{color}

\begin{document}
\title{Addressing Situated Teaching Needs: A Multi-Agent Framework for Automated Slide Adaptation}
\titlerunning{Addressing Situated Teaching Needs}
\author{Binglin Liu \and
Yucheng Wang \and
Zheyuan Zhang \and
Jiyuan Lu \and
Shen Yang \and
Daniel Zhang-Li \and
Huiqin Liu \and
Jifan Yu\thanks{Corresponding author.}
}
\authorrunning{B. Liu et al.}
\institute{Tsinghua University, Beijing, China \\
\email{lbl23@mails.tsinghua.edu.cn, yujifan@tsinghua.edu.cn}}
\maketitle              
\begin{abstract}
The adaptation of teaching slides to instructors' situated teaching needs, including pedagogical styles and their students' context, is a critical yet time-consuming task for educators. Through a series of educator interviews, we first identify and systematically categorize the key friction points that impede this adaptation process. Grounded in these findings, we introduce a novel multi-agent framework designed to automate slide adaptation based on high-level instructor specifications. 
An evaluation involving 16 modification requests across 8 real-world courses validates our approach. The framework's output consistently achieved high scores in intent alignment, content coherence and factual accuracy, and performed on par with baseline methods regarding visual clarity, while also demonstrating appropriate timeliness and a high operational agreement with human experts, achieving an F1 score of 0.89. This work heralds a new paradigm where AI agents handle the logistical burdens of instructional design, liberating educators to focus on the creative and strategic aspects of teaching.

\end{abstract}
\section{Introduction}
Presentation slides have become a cornerstone of modern educational practice, serving as a primary and effective medium for delivering instructional content across various academic disciplines~\cite{daniels1999introducing,strauss2011optimizing}. Yet, their full pedagogical value is often unlocked not at the moment of creation, but through the process of constant adaptation.

As one art history lecturer noted in our formative study, adapting an existing slide deck originally designed for majors to suit a general education course requires fundamental changes: for majors, the presentation emphasizes technical analysis, while for non-majors, it leans into narrative. This scenario illustrates one form of what we term \textit{slide adaptation}. More broadly, slide adaptation is the systematic process of modifying a slide deck's content, structure, and multi-modal elements to address situated teaching needs, encompassing instructors' pedagogical styles, students' cognitive levels, curriculum standards, related heated topics, \textit{etc}. The ultimate goal is to achieve alignment with teachers' curriculum vision, thereby maximizing instructional effectiveness~\cite{choppin2011learned}.

This adaptive work is critical for effective teaching, yet it represents a significant and often unacknowledged investment of an instructor's time. The recent advent of AI-powered slide generation tools, such as Manus\footnote{\url{https://manus.im/playbook/slide-generator}} and Tongyi PPT\footnote{\url{https://www.aitags.cn/}}, alongside research utilizing Large Language Models (LLMs) and agent systems~\cite{mondal2024presentations,shenoy2025automated,zheng2025pptagent,ge2025autopresent}, offers new potential to elevated efficiency within the slide adaptation workflow. However, a significant disconnect exists, rooted in two primary limitations. First, these platforms are designed for content generation rather than refinement and adaptation. Second, the quality of their output often fails to meet the professional and pedagogical standards required for effective instruction. While recent works like Auto-Slides~\cite{yang2025auto} have made significant strides in automated slide modification, their focus remains self-learning-supportive and document-driven, neglecting the instructor's perspective. This reveals a critical gap in the realm of teaching slide adaptation.

To address these gaps, we propose and explore the following research questions.
\begin{itemize}
\item \textbf{RQ1}: What are the primary challenges in the manual workflow of slide adaptation for instructors?
\item \textbf{RQ2}: How can a multi-agent framework be designed to translate teachers' pedagogical intentions into concrete slide modifications?
\item \textbf{RQ3}: How does the proposed framework's performance in slide adaptation efficiency and quality compare against the manual process and existing AI-powered solutions?
\end{itemize}

To answer these questions, we first conducted semi-structured interviews with teachers from various fields to investigate \textbf{RQ1}. Informed by the findings, we then designed and implemented a multi-agent prototype system to explore \textbf{RQ2}. This system interprets and executes an instructor's adaptation needs for a given set of slides, ensuring efficiency without sacrificing pedagogical control. We then evaluated the system by measuring task completion time and the generation quality of generated slides to answer \textbf{RQ3}. 

The experimental evaluation was conducted on a selection of cases. To assess efficiency, we compared the modification time of our system against baseline approaches. For quality assessment, we conducted an operation-based evaluation and developed a rubric on outputs comprising four dimensions: \textit{intent alignment}, \textit{content coherence}, \textit{factual accuracy}, and \textit{visual clarity}.

The core contribution of this work is \textbf{a distinctive multi-agent framework for automated teaching slide adaptation that helps translate an educator's natural language intent into concrete modification actions}, offering a systematic approach to tackle the complexities inherent in situated teaching needs. Our key technical innovation is a hierarchical architecture that effectively decouples high-level curriculum contextualization from the fine-grained execution of content changes. This layered design, by establishing distinct focuses across different stages, streamlines the customization process, empowering educators to efficiently tailor instructional materials to their precise specifications. The framework's utility spans the full content lifecycle, facilitating both the annual refinement of slides and the seamless adaptation of Open Educational Resources (OER) for specific classroom contexts. Reducing educators' logistical burdens, this work fundamentally empowers the evolving needs of teaching and learning.

\section{Preliminaries}
\subsection{Related Work}
\subsubsection{Automated Slide Generation \& Adaptation}
The field of automated slide generation has evolved from early extractive and semantic methods~\cite{utiyama1999automatic,yasumura2003support} to sophisticated neural and multimodal systems. Neural network-based models, such as PPSGen~\cite{hu2013ppsgen}, DOC2PPT~\cite{fu2022doc2ppt} and SlideGen~\cite{sefid2021slidegen}, achieve superior content and structural coherence compared to traditional summarizers. More recently, LLM-based systems have opened up possibilities for audience-targeted~\cite{mondal2024presentations} and reference-aligned~\cite{zheng2025pptagent} presentation generation. Despite these advancements, most existing frameworks are limited to one-shot generation, lacking capabilities for iterative refinement.

Regarding slide adaptation, a recent approach gathers valuable guidance to presenters~\cite{warner2023slidespecs}, but the process fall short of automation. A significant step forward was made by Auto-Slides~\cite{yang2025auto} and AutoPresent~\cite{ge2025autopresent}, which achieved automated generation and modification, emphasizing iterative refinement. However, existing systems have predominantly been applied to tasks such as learning from academic documents~\cite{yang2025auto} and generic natural language-to-PPT conversion~\cite{ge2025autopresent}. Consequently, teacher-centric methodologies remain a significant and underexplored area of research.

This paper aims to address this gap by focusing on adaptive and teaching-centered slide modification.

\subsubsection{AI-assisted Lesson Planning}
Lesson planning is a core element of effective teaching~\cite{farhanglesson} and plays a central role in slide adaptation. Early explorations highlighted its potential to streamline lesson preparation~\cite{belloula2025empowering,vitanova2025utilization}, with subsequent studies reporting positive outcomes in automated lesson preparing~\cite{karaman2024lesson}. Moreover, pre-service teachers can benefit from AI collaboration as a form of professional development \cite{sun2025exploring}. 
Building on this trajectory, researchers have designed agents and planning tools to extend teachers' pedagogical capacity~\cite{baylor2002expanding,dennison2025teacher,zhang2025eduplanner}. However, concerns persist that these agents may lack the specificity to generate novel ideas for seasoned teachers and ``have minimal alignment with'' established planning frameworks~\cite{lammert2024better}.

This work further investigates methods to foster effective collaboration and alignment between teachers and AI, specifically within the automated planning module for slide adaptation.

\subsection{Problem Formulation}

We formulate the core process of slide adaptation as a transformation where an intelligent agent, $\mathcal{A}$, revises a presentation based on a natural language request and prepared materials from an educator. An initial presentation is represented by a state $S = \{s_1, \dots, s_N\}$, where each slide $s_i$ is a structured tuple $(T_i, I_i, L_i)$ comprising its textual content, imagery, and layout. The agent $\mathcal{A}$ takes the initial presentation $S$ and the educator's specific request $R$ alongside the materials $M$ as input to generate a revised presentation, $S'$. This state transition is defined as:
$$S' = \mathcal{A}(S, R, M)$$

The central objective is to design the agent $\mathcal{A}$ such that the revised presentation $S'$ satisfies the situated pedagogical needs. While the full adaptation process ranges from requirement formation to confirmation and fine-tuning, our study distills its essential unit, an approach that remains extensible.

\section{Formative Studies}
To investigate the common practice and perceived challenges in slide adaptation and inform our system design, we conducted semi-structured interviews (N = 5) with educators spanning primary, secondary, and tertiary education levels.

\subsection{Participants and Procedure}
We recruited five educators (3 female, denoted P1-P5) through institutional partnerships and professional networks to ensure diversity in teaching experience and educational levels (see Table \ref{tab:participants}). A key inclusion criterion was that all participants frequently used slides as a teaching tool. Participants had an average of 17.8 years of teaching experience (SD = 9.31).

Interviews were conducted via video conference, each lasting approximately 30 minutes. The interview protocol explored three core domains:
(1) current slide development workflows and challenges, (2) personal experiences of AI-assisted slide adaptation, and (3) desired features for future automation tools.
All interviews were audio-recorded and transcribed verbatim for analysis.

\begin{table}[htbp]
    \caption{Participant Demographics (Exp means Years of Teaching Experience)}
    \label{tab:participants}
    \centering
    \begin{tabular}{lllll}
        \toprule
        \textbf{ID} & \textbf{Level} & \textbf{Subject} & \textbf{Exp} & \textbf{Gender} \\
        \midrule
        P1 & Secondary & Biology & 21 & Female \\ 
        P2 & Tertiary & Chemical Education & 30 & Male \\ 
        P3 & Primary & Mathematics & 10 & Female \\ 
        P4 & Tertiary & Art History & 7 & Male \\ 
        P5 & Tertiary & Enterprise Management & 21 & Female \\ 
        \bottomrule
    \end{tabular}
\end{table}

\subsection{Data Analysis}
We performed a thematic analysis~\cite{braun2006using} of the interview transcripts. Two researchers independently developed an initial codebook from the first two transcripts, resolving discrepancies through discussion. The remaining transcripts were coded by one researcher and reviewed by the second for consistency. Axial coding was then used to group the codes into higher-level themes corresponding to our research questions.

\subsection{Findings}
Our analysis identified a consistent set of practices, challenges, and expectations surrounding the slide adaptation process. We synthesize these insights into four key findings.
\begin{itemize}
    \item \textbf{F1: The Adaptation Workflow is a Laborious and Predominantly Manual Endeavor.} Participants' adaptation patterns varied. P1 described ``continuously adding and modifying content'', a practice shared with P4, where both primarily made minor revisions. In contrast, P2 advocated for major overhauls to align with evolving practices in chemistry education. Despite these variations, the adaptation process consistently addressed \textit{three dimensions}: (1) visualization (aesthetics, layout); (2) content (adjusting examples, language); and (3) narrative structure (re-sequencing slides). Regarding generative AI (GenAI), participants primarily leveraged it for tasks such as simulating student responses (P1), organizing materials and expressions (P2), and generating interactive presentation components (P3). They rarely, however, attempted to use AI for direct slide modification. This is notable, as manually revising a full lesson's slides was reported to take over one hour by P1, P4 and P5.
    \item \textbf{F2: Adaptation is Driven by Situated Teaching Needs.} The primary impetus for adaptation is the alignment of teaching slides with specific, situated needs. P3 defined such needs as twofold: ``my own teaching styles and the students' learning situation''. In detail, these \textit{situated} factors, both within and beyond the classroom, may include the school's ``requirements for class activity arrangement'' (P1), changes in frontline industry practices (P2), the receptivity of students in different classes (P3), and ``currently popular'' contents (P4). Successfully navigating these factors is central to effective adaptation.
    \item \textbf{F3: The Core Challenges Lie in Content Processing and Narrative Structuring.} The most significant challenges reported by participants were concentrated within two dimensions: content (P1, P2, P3, P5) and narrative structure (P4). P1 described sourcing appropriate materials as ``quite painful'', while P5 stressed the challenge of fact-checking. Distilling complex information into a simplified, structured whole was also mentioned by P2. Regarding the narrative structure dimension, P4 explicitly identified “building the framework” as the most difficult part, pinpointing the challenge of organizing the lesson flow and logic. These areas were consequently highlighted as having the greatest potential for beneficial computational assistance.
    \item \textbf{F4: Apprehension Over Quality and Applicability.} Participants were concerned about the quality of the generated content itself (\textit{i.e.}, its correctness and clarity of presentation, P2, P3, P4, P5). This apprehension wasn't about rejecting AI, but about demanding a more transparent and controllable collaborative model. Educators need the ability to readily scrutinize and revise the outputs, which is the key to safeguarding quality and pedagogical soundness.
\end{itemize}

Synthesizing our findings, we derived three design goals (\textbf{DG}) that form a blueprint for a system that is effective, efficient, and teacher-friendly.
\begin{itemize}
    \item \textbf{DG1}: Prioritize Context-Aware, High-Fidelity Generation. (F2, F3)
    \item \textbf{DG2}: Provide Streamlined and Integrated Support across Three Dimensions. (F1)
    \item \textbf{DG3} Uphold Educator Agency through Controllable, Explainable AI. (F4)
\end{itemize}

\section{System Design}
\subsection{System Overview and Architecture}
\begin{figure}[htbp]
    \centering
    \includegraphics[width=0.7\columnwidth]{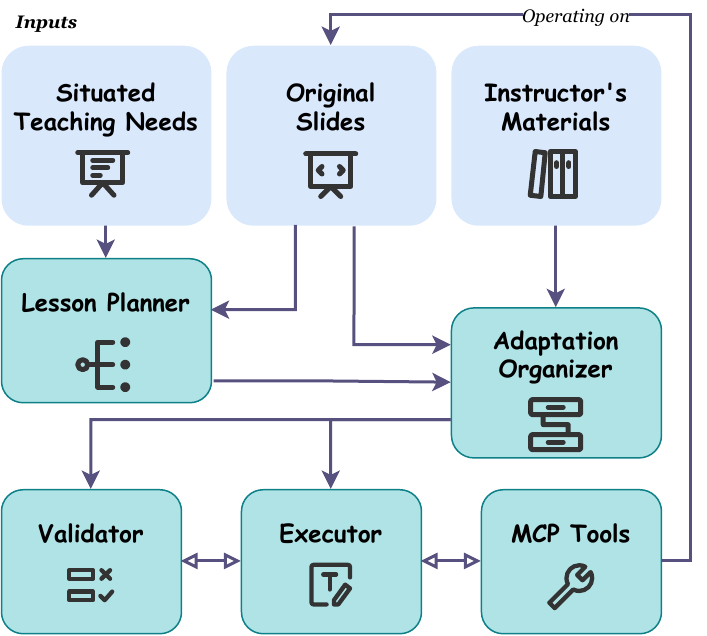} 
    \caption{Framework overview.}
    \label{fig:framework}
\end{figure}

This section details our proposed Multi-Agent Framework (\Framework), as illustrated in Figure \ref{fig:framework}. The architecture is organized into two primary phases: a \textit{Strategic Planning} phase, where the Lesson Planner (\textbf{LP}) and Adaptation Organizer (\textbf{AO}) collaboratively generate a fine-grained adaptation plan, and an \textit{Execution and Quality Assurance} phase, where the Executor (\textbf{EX}) applies the changes and the Validator (\textbf{VA}) verifies the outcome against the initial objectives.

The operational architecture of \Framework is predicated on three primary inputs: Situated Teaching Needs (\textbf{STN}), which define the high-level pedagogical objectives; Original Slides (\textbf{OS}), which serve as the source material for modification; and Instructor's Materials (\textbf{IM}), which provide reference contents for generation.

\subsection{Phase 1: Strategic Planning and Organization}
This initial phase is structured as a hierarchical generation process, translating an educator's terse request into a concrete, actionable plan.  The system's first step is to categorize the educator's situated teaching needs into two distinct types, which dictates the subsequent workflow: \textit{Recompose} for comprehensive structural overhauls, and \textit{Refine} for targeted, content-focused modifications.

\textbf{Lesson Planner (LP)}: The Lesson Planner's role adapts based on the initial classification.  This bifurcated approach ensures both pedagogical robustness for major revisions and operational efficiency for minor tweaks.

For \textit{Recompose} requests, the Lesson Planner ingests the educator's request and the original slides to synthesize a high-level pedagogical framework. To achieve context-aware generation (\textbf{DG1}), it applies pedagogical principles to analyze the teaching needs, formulating a comprehensive \textit{instructional design}, which specifies pedagogical elements comprising basic course information, student analysis, instructional flow of the lesson, learning objectives, and difficult points for students. By performing a gap analysis between this ideal design and the existing slides, the \textbf{LP} then generates a high-level \textit{adaptive guideline}. This guideline serves as an explainable pedagogical arrangements (\textbf{DG3}), outlining the core modifications required and making the AI's high-level reasoning transparent to the educator.

For \textit{Refine} requests, which focus on minor content adjustments, the \textbf{LP} adopts a more direct and efficient approach. It parses the educator's instruction and translate it into a precise, actionable directive.

\textbf{Adaptation Organizer (AO)}: Functioning as the system's strategist, the Adaptation Organizer transforms the directives from Lesson Planner, cross-referencing it with the \textbf{OS} and \textbf{IM}, to a concrete, atomic sequence of operations for each slide. This directly supports streamlined and integrated support (\textbf{DG2}). 

The adaptation organizing process begins with resource augmentation. Guided by the instructional design, the \textbf{AO} formulates contextual queries tailored to each topic presented in the lesson and utilizes the Tavily Search API\footnote{\url{https://app.tavily.com/}} to retrieve a curated information packet, including up-to-date text and available image URLs (\textbf{DG1}). Subsequently, the retrieved resources and other materials are synthesized into a structural mapping---defining which slides to ADD, MODIFY, or KEEP---with granular execution instructions for each page, which provides a controllable and explainable manifest (\textbf{DG3}) of all intended changes before execution.

\begin{figure}[htbp]
    \centering
    \includegraphics[width=0.82\columnwidth]{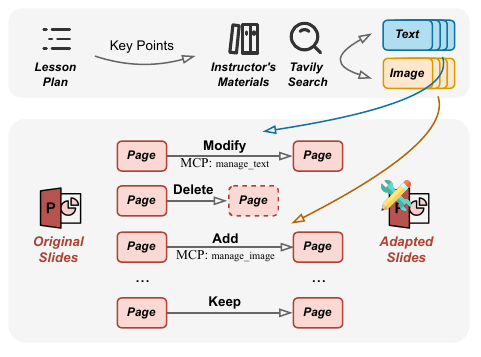} 
    \caption{Workflow of the Adaptation Organizer. Mappings from Original Slides to Adapted Slides and detailed JSON-formed execution steps are established at this stage.}
    \label{fig:workflow}
\end{figure}

\subsection{Phase 2: Execution and Quality Assurance}
The execution of the plan and the verification of its outcome are managed by two agents:

\textbf{Executor (EX)}: The Executor serves as the core operational agent within our framework, parsing and executing the instruction sequence from the \textbf{AO} by interfacing with a custom-designed set of MCP (Model Context Protocol) tools. Utilizing the python-pptx project\footnote{\url{https://pypi.org/project/python-pptx/}}, a robust and widely-adopted Python library for programmatic interaction with PowerPoint files, we engineered each tool to correspond directly to a specific atomic operation (see Table \ref{tab:MCPTools}). 

\begin{table}[htbp]
\centering
\caption{Description of MCP tools and applicable scenarios.}
\label{tab:MCPTools}
\begin{tabularx}{\linewidth}{
  >{\raggedright\arraybackslash}p{2.7cm}
  X
}
\toprule
\textbf{Scenario} & \textbf{Tool and Description} \\
\midrule
\multirow{2}{*}{\centering Modify Only} 
    & \texttt{extract\_slide\_text}: Extracts all text from text boxes and placeholders within a single slide. \\[3pt]
    & \texttt{get\_slide\_images\_info\_for\_llm}: Sends slide images to the LLM for semantic understanding. \\ 
\midrule
\multirow{2}{*}{\centering Add \& Modify} 
    & \texttt{manage\_text}: 
      (1) Deletes or replaces text within a text box. 
      (2) Adds text boxes with customizable position, size, and fonts. \\[3pt]
    & \texttt{manage\_image}: 
      (1) Deletes or replaces images by ID. 
      (2) Adds images from URLs, base64, or local paths, with adjustable position, scale, and size. \\
\bottomrule
\end{tabularx}
\end{table}

\textbf{Validator (VA)}: The Validator functions as the quality assurance module, guaranteeing high-fidelity generation (\textbf{DG1}). Following each execution, it assesses the modified slides against the execution instructions on a per-slide basis. Through content and typographical analysis, it identifies discrepancies, which are then rectified by the Executor in an iterative feedback loop. This operational cycle concludes upon successful validation, yielding the final output.

\section{Experiments}
\subsection{Metrics}
Grounded in established theories such as Mayer's Cognitive Theory of Multimedia Learning~\cite{mayer2002multimedia} and Sweller's Cognitive Load Theory~\cite{sweller1988cognitive} and informed by our \textbf{DG}s, we designed the following metrics to evaluate the system's output:

\begin{enumerate}
    \item \textbf{Intent Alignment (IA)}: How well the generated adaptations reflect the educator's specified needs.
    \item \textbf{Content Coherence (CC)}: The logical consistency, clarity, and narrative flow of the adapted material.
    \item \textbf{Factual Accuracy (FA)}: The correctness and veracity of any newly introduced or modified information.
    \item \textbf{Visual Clarity (VC)}: The legibility, comprehensibility, and pedagogical effectiveness of the visual design.
\end{enumerate}
The detailed evaluation rubrics for each metric are provided in Appendix A.

\subsection{Experimental Setup}
Our evaluation is designed to assess the framework's performance across a diverse set of real-world teaching scenarios.
We curated a dataset of 8 slide decks (10-25 slides each) from OER platforms, spanning computer science, business, and humanities. We formulated 2 distinct modification requests for each slide deck, creating 8 recompose and 8 refine tasks as representative of common situated teaching needs, ranging from contextualizing for a different audience, updating case studies with hot spot issues, or incorporating newly-prepared materials, to redesigning a dense slide into an interactive activity.

\subsection{Evaluation Protocol}
We evaluated outputs from three systems: our proposed system and two baselines, GLM Slide/Poster Agent\footnote{\url{https://docs.z.ai/guides/agents/slide}} and Office-PowerPoint-MCP-Server\footnote{\url{https://github.com/GongRzhe/Office-PowerPoint-MCP-Server}}. Outputs were assessed using a dual methodology (human experts and LLMs).

\subsubsection{Operational Evaluation.} To create a ground-truth for operational evaluation, three pedagogical experts co-built a reference set of necessary and optimal operations for 5 selected tasks. Crucially, to avoid the subjective evaluation of creative adaptations, the selected tasks were designed with explicit goals, and our analysis focuses solely on the execution of the specified operational sequence. We then quantified each system's performance by comparing its executed operations against this reference set. This comparison was framed by identifying correctly executed actions (True Positives), omissions (False Negatives), and deviations/extraneous actions (False Positives) to calculate precision, recall, and F1-score.

\subsubsection{Output Evaluation.} In addition to the operational analysis, We selected the Gemini-2.5-Pro model~\cite{comanici2025gemini} to build an automated evaluation pipeline of \textit{final output} designed to supplant human judgment. This model was chosen as a representative benchmark for the current state-of-the-art in ``LLM-as-a-judge'' evaluation. To ensure deterministic and reproducible results, the model's temperature was set to 0. To ensure a fair comparison, the LLM was tasked with emulating the human evaluation protocol. For each evaluation instance, the model's inputs consisted of the original slides, the instructor's request, the corresponding system-generated output, and the evaluation rubrics. The LLM was prompted to produce a quantitative score (1-5) and a qualitative justification for each of the rubrics, replicating the dual-scoring (score and rationale) methodology of the human experts.

\subsection{Results}
\subsubsection{Operational Evaluation Results}
We evaluated each system's generated operations ($\mathcal{S}$) against an expert-defined reference set ($\mathcal{R}$) using precision, recall, and F1-score, defined as follows:
\begin{itemize}
    \item \textbf{True Positives (TP)}: Operations in $|\mathcal{S} \cap \mathcal{R}|$.
    \item \textbf{False Positives (FP)}: Operations in $|\mathcal{S} - \mathcal{R}|$.
    \item \textbf{False Negatives (FN)}: Operations in $|\mathcal{R} - \mathcal{S}|$.
\end{itemize}

As shown in Table~\ref{tab:ground_truth_ops}, our system demonstrates superior performance, achieving an F1-Score of $0.89$. This score is underpinned by a strong balance between high precision ($P=0.90, R=0.88$). The high precision signifies that the executed actions are highly relevant to user intent, minimizing superfluous operations, while the high recall ensures the comprehensive execution of nearly all required steps, thus avoiding critical omissions.

\begin{table}[htbp]
\centering
\caption{Operational performance comparison based on the ground-truth evaluation. \textbf{Bolded}
 represent the largest value.}
\label{tab:ground_truth_ops}
\begin{tabular}{lccc}
\toprule
\multicolumn{1}{c}{\textbf{Approach}} & \textbf{Precision} & \textbf{Recall} & \textbf{F1-Score} \\
\midrule
GLM  & 0.73    & 0.83    & 0.76  \\
PPT-MCP  & 0.31  & 0.18   & 0.20  \\
\textbf{Ours} & \textbf{0.90} & \textbf{0.88}  & \textbf{0.89}  \\
\bottomrule
\end{tabular}
\end{table}

In contrast, the baselines demonstrate pronounced deficiencies. The GLM approach, despite a moderate F1-Score of $0.76$, exhibits a significant precision-recall imbalance ($P=0.73, R=0.83$). This low precision is primarily attributed to its tendency towards over-generation, performing unnecessary modifications such as replacing existing images and text beyond the user's explicit request.

The PPT-MCP approach performs worse, with an F1-Score of $0.20$, which stems from a critical deficiency in instruction following, particularly in maintaining accuracy over long contexts. This is compounded by a limited capacity for understanding and breaking down the instructions, leading to a fundamental failure to identify the required task sequence.

In summary, this quantitative analysis against ground truth underscores the capability of our approach in deconstructing and accurately executing complex and general slide modification instructions.

\subsubsection{Overall Performance Evaluation Based on Output}
As shown in Table~\ref{tab:main_results}, our proposed system achieves substantial performance gains over both baselines across the majority of metrics except Time and Visual Clarity.

\begin{table}[htbp]
\centering
\caption{LLM-Assessed Evaluation Results. GLM and PPT-MCP respectively stands for GLM Slide/Poster Agent and Office-PowerPoint-MCP-Server. \textbf{Bolded} and \underline{underlined} represent the first and
second largest value.}
\label{tab:main_results}
\begin{tabular*}{\columnwidth}{@{\extracolsep{\fill}} lccccc}
\toprule
\textbf{Approach} & \textbf{T(s)} & \textbf{IA} & \textbf{CC} & \textbf{FA} & \textbf{VC} \\
\midrule
GLM & 330 & \underline{3.56} & \underline{2.19} & 1.50 & \underline{3.22} \\
PPT-MCP & \textbf{17} & 2.53 & 1.88 & \underline{2.13} & \textbf{3.31} \\
\textbf{Ours} & \underline{159} & \textbf{3.75} & \textbf{2.78} & \textbf{2.38} & 3.16 \\
\bottomrule
\end{tabular*}
\end{table}

Our framework achieved the highest score in Intent Alignment, indicating that our system is substantially more effective at fulfilling the educator's explicit directives and avoiding superfluous content. This success is primarily attributed to our multi-agent architecture, where dedicated planning and reflection agents ensure a more thorough and goal-oriented interpretation of high-level specifications.

Our approach surpassed both baselines in Content Coherence, nearing the ``Adequate'' threshold and suggesting that generated slides generally maintain a more logical flow and internal consistency. The framework also showed improved performance in Factual Accuracy, mitigating hallucination and underscoring its ability to incorporate factual grounding better than baselines. Despite improvements, these metrics reveal the framework is constrained by residual structural weaknesses and a persistent insufficiency in factual accuracy, the latter of which we will explore in the next section.

Regarding Visual Clarity, our framework performed on par with the baseline methods. While it ensures better stylistic consistency with the original deck, the process of editing the original design may compromise legibility and comprehensibility. This outcome indicates that our framework prioritizes the logical and semantic aspects of slide adaptation over fine-grained visual and stylistic manipulation, resulting in a competent but not visually superior output.

\begin{figure}[htbp]
    \centering
    \includegraphics[width=0.85\columnwidth]{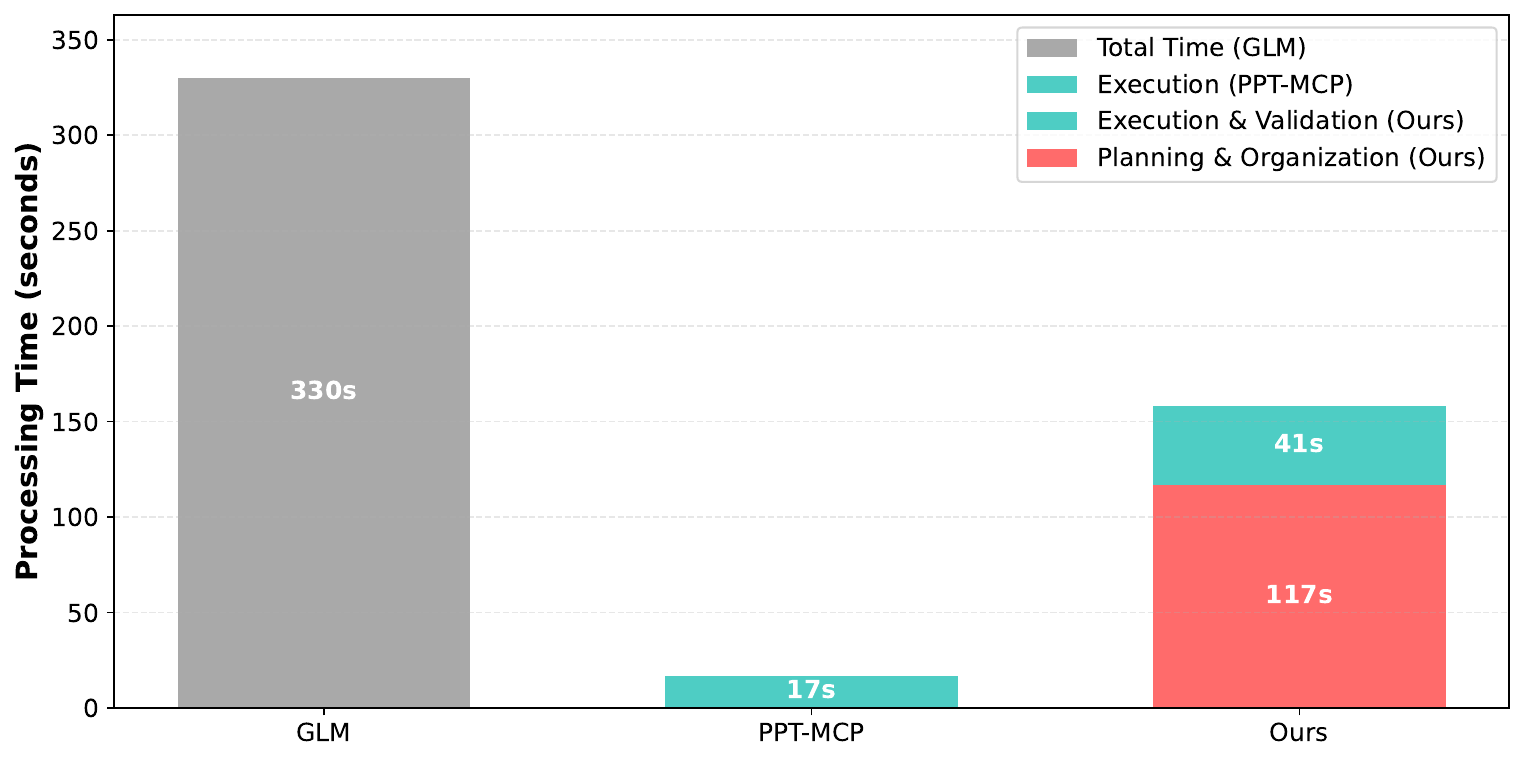} 
    \caption{Processing time comparison across methods.}
    \label{fig:time}
\end{figure}

In terms of efficiency, our framework is faster than GLM (via parallel processing) but slower than PPT-MCP, highlighting a speed-quality trade-off. PPT-MCP sacrifices quality for speed, whereas our multi-step process ensures higher performance at the cost of longer processing, but still within an acceptable waiting window for users.

In summary, our multi-agent framework validates a promising exploratory step toward automated slide adaptation, demonstrating a clear advantage over existing methods in aligning with instructor intent and improving content coherence and factual accuracy. Nevertheless, areas such as visual clarity and overall performance indicate a need for further refinement, positioning the system as a powerful first-step assistant rather than a standalone solution.

\section{Discussion and Future Works}
\subsubsection{Architectural Paradigms: Regeneration vs. In-situ Modification}
A key distinction exists between two technical approaches for modifying lecture slides: regeneration and in-situ modification. Many AI-empowered slide tools, including Gamma\footnote{\url{https://gamma.app/en/products/presentations}} and GLM, regenerate the entire slide as new code (\textit{e.g.}, HTML/JSON), a process that often overwrites a teacher's established visual style and embedded images. While this can yield aesthetically pleasing results, it fails to respect the pedagogical context embedded in the original design. Our approach, in contrast, directly modifies the existing slide. This in-situ approach preserves the original design language, which is crucial for maintaining consistency across course materials; meanwhile, the visual effects might suffer. This architectural choice helps to explain our modest performance in Visual Clarity.

\subsubsection{Plan-Act Framework in Slide Adaptation}
Our findings provide strong empirical validation for the enhanced intent alignment within the Plan-Act paradigm~\cite{erdoganplan,zhu2025knowagent}, specifically within the context of slide modification. In our framework, the intermediate plan drafted by Adaptation Organizer acts as a crucial blueprint, ensuring the subsequent actions are logical, sequential, and directly tied to the educator's goal. The stark performance contrast with the PPT-MCP baseline underscores this point. Lacking an explicit planning stage, it often fails to correctly interpret the sequence or execute the right operations. Our results demonstrate that for the nuanced task of slide modification, introducing an explicit planning step is essential. It translates the teacher's abstract goal into a concrete, executable workflow, leading to a significant performance improvement over direct-action models.

\subsubsection{The Critical Gap in Factual Accuracy}
The low Factual Accuracy scores across all systems, including web-search-augmented approaches (ours \& GLM), highlight a critical and persistent challenge in LLM-based content generation: the phenomenon of hallucination. Consistent with findings in the literature~\cite{ji2023survey}, relying on Retrieval-Augmented Generation (RAG) is not a panacea; models frequently struggle with context misinterpretation or knowledge override when synthesizing information from retrieved snippets, while the content quality from searches is fluctuating. The limited and varying factual reliability of our approach (FA: $Avg=2.38, SD=1.45$) further confirms this position and necessitates post-generation human validation to ensure accuracy in educational materials.

\subsubsection{Limitations and Envisions}
While our approach demonstrates strong performance on the evaluated tasks, this work also reveals several key limitations that chart the course for future development.

First, our evaluation was confined to presentations of a maximum of 25 pages, a scope that does not fully capture the complexity of long-term, real-world scenarios such as semester-long lecture series. We envision future work focusing on scaling the system's context management capabilities, potentially through hierarchical memory mechanisms, to ensure robust performance across both more extended and more varied environments.

Second, support for rich multi-modal resources, such as embedded videos, complex charts, and diagrams, is limited. We envision integrating multi-modal foundation models to enable the agent to not only ``see'' but also ``understand'' and manipulate visual components, moving towards a truly holistic understanding and control of presentation content.

Moreover, our current system focuses on single-turn interaction. This design choice was deliberate, aimed at simplifying the problem space to isolate and validate the core challenge of slide adaptation. However, real-world slide adaptation is often an iterative process. We envision a key direction for future work to be enabling multi-turn, iterative refinement, shifting the paradigm from command execution to collaborative teaching design.

\section{Conclusion}
This study investigated the specialized task of teaching slide adaptation, identifying content processing and narrative structuring as key challenges faced by educators. To address these challenges, we proposed and implemented a novel multi-agent, plan-act framework designed to automate the revision of teaching materials based on situated teaching needs. Validation using real-world adaptation requests demonstrated that the proposed system achieves high operational agreement with experts and exhibits strong performance across key evaluative metrics. Ultimately, this research establishes the viability of utilizing multi-agent AI architectures to manage the logistical complexities of slide revision and refinement, thereby liberating educators to focus on the higher-order and creative dimensions of their pedagogical practice.

\begin{credits}
\subsubsection{\ackname} This work was supported by the National Natural Science Foundation of China (62407027), Tsinghua  University Initiative Scientific Research Program (2024THZWJC11), and a grant of SMP-Zhipu20240211.

\subsubsection{\discintname}
The authors have no competing interests to declare that are relevant to the content of this article.
\end{credits}
%
\bibliographystyle{splncs04}
\bibliography{ref}

\appendix
\section{Evaluation Rubrics}
The evaluation of adapted slides are conducted using the following criteria, each assessed on a 5-point Likert scale (5 = Excellent, 4 = Very Good, 3 = Adequate, 2 = Needs Improvement, 1 = Poor).

\begin{itemize}
    \item \textbf{Intent Alignment (IA)}
    \begin{itemize}
        \item \textbf{IA-1: Fulfillment of Explicit Directives} Evaluates the extent to which the generated output comprehensively and accurately addresses all specified requirements and constraints from the educator, ensuring no omissions or misinterpretations of the given tasks.
        \item \textbf{IA-2: Avoidance of Superfluous Content} Assesses whether the system introduced any extraneous or unsolicited information that deviates from the educator's original intent.
    \end{itemize}

    \item \textbf{Content Coherence (CC)}
    \begin{itemize}
        \item \textbf{CC-1: Logical Consistency} Measures the internal integrity of the material, focusing on whether ideas, arguments, and concepts are presented in a rational, non-contradictory sequence.
        \item \textbf{CC-2: Narrative Flow} Evaluates the structural quality of the content, ensuring a well-structured progression of ideas so that the instructional message is communicated effectively and without ambiguity.
    \end{itemize}

    \item \textbf{Factual Accuracy (FA)}
    \begin{itemize}
        \item \textbf{FA-1: Verifiability and Correctness} Scrutinizes all informational content (new facts, modified data, supplemental examples) to ensure they are factually sound, current, and verifiable.
    \end{itemize}

    \item \textbf{Visual Clarity (VC)}
    \begin{itemize}
        \item \textbf{VC-1: Legibility and Comprehensibility} Assesses that the typography, layout, and graphical elements are clear, easily understandable, and pedagogically effective in facilitating learning.
        \item \textbf{VC-2: Stylistic Consistency} Measures the congruence of the visual design of the adapted slides with the original slides, ensuring a seamless user experience.
    \end{itemize}
\end{itemize}

\subsubsection{Scoring Detail (5-Point Likert Scale)}

\begin{itemize}
    \item \textbf{5 (Excellent):} Criterion is met completely with exceptional quality and no discernible flaws.
    \item \textbf{4 (Very Good):} Criterion is met very well; only minor, non-disruptive flaws or inconsistencies are present.
    \item \textbf{3 (Adequate):} Criterion is acceptably met; some flaws are present, but they do not significantly impede the pedagogical goal.
    \item \textbf{2 (Needs Improvement):} Criterion is only partially met; significant flaws or inconsistencies that impede effective learning are present.
    \item \textbf{1 (Poor):} Criterion is not met; fundamental failures in the output render it unusable or counterproductive for the intended pedagogical purpose.
\end{itemize}

\end{document}